\documentclass[pdflatex,sn-mathphys-num]{sn-jnl}

\usepackage{diagbox}
\usepackage{graphicx}%
\usepackage{multirow}%
\usepackage{amsmath,amssymb,amsfonts}%
\usepackage{amsthm}%
\usepackage{mathrsfs}%
\usepackage[title]{appendix}%
\usepackage{xcolor}%
\usepackage{textcomp}%
\usepackage{manyfoot}%
\usepackage{booktabs}%
\usepackage{algorithm}%
\usepackage{algorithmicx}%
\usepackage{algpseudocode}%
\usepackage{listings}%

\theoremstyle{thmstyleone}%
\theoremstyle{thmstyletwo}%

\theoremstyle{thmstylethree}%

\begin{document}

\title[Article Title]{Programmable Photonic Extreme Learning Machines}


\author*[1]{\fnm{José Roberto} \sur{Rausell-Campo}}\email{joraucam@upv.es}

\author[2]{\fnm{Antonio} \sur{Hurtado}}\email{antonio.hurtado@strath.ac.uk}

\author[3]{\fnm{Daniel} \sur{Pérez-López}}\email{daniel.perez@ipronics.com}

\author[1]{\fnm{José} \sur{Capmany Francoy}}\email{jcapmany@iteam.upv.es}

\affil*[1]{\orgdiv{Photonics Research Lab, ITEAM}, \orgname{Universitat Politècnica de Valencia}, \orgaddress{\state{Valencia}, \country{Spain}}}

\affil[2]{\orgdiv{Institute of Photonics, Physics Department}, \orgname{University of Strathclyde}, \orgaddress{\state{Glasgow}, \country{United Kingdom}}}

\affil[3]{\orgdiv{iPronics Programmable Photonics S.L.}, \orgaddress{\state{Valencia}, \country{Spain}}}


\abstract{Photonic neural networks offer a promising alternative to traditional electronic systems for machine learning accelerators due to their low latency and energy efficiency. However, the challenge of implementing the backpropagation algorithm during training has limited their development. To address this, alternative machine learning schemes, such as extreme learning machines (ELMs), have been proposed. ELMs use a random hidden layer to increase the feature space dimensionality, requiring only the output layer to be trained through linear regression, thus reducing training complexity. Here, we experimentally demonstrate a programmable photonic extreme learning machine (PPELM) using a hexagonal waveguide mesh, and which enables to program directly on chip the input feature vector and the random hidden layer. our system also permits to apply the nonlinearity directly on-chip by using the system's integrated photodetecting elements. Using the PPELM we solved successfully three different complex classification tasks. Additioanlly, we also propose and demonstrate two techniques to increase the accuracy of the models and reduce their variability using an evolutionary algorithm and a wavelength division multiplexing approach, obtaining excellent performance. Our results show that programmable photonic processors may become a feasible way to train competitive machine learning models on a versatile and compact platform.}





\maketitle

\section{Introduction}\label{intro}

In recent years, artificial intelligence algorithms have gained incresing attention due to the development of more powerful and complex architectures \cite{NIPS2017_3f5ee243,10.5555/3495724.3496298}.  This exponential increase in the size and number of required parameters has led to a higher demand for computational resources. Traditional electronic processors have been the backbone of these advancements, but they present limitations in bandwidth and energy efficiency. Consequently, alternative computing paradigms that address these issues have started to attract increasing research 
attention\cite{Stark2020,Tait2017,Shastri2021,Liao2023}.

In this context, photonics has emerged as a viable alternative for machine learning hardware by exploiting the unique features of light-based computing, including complex-valued, multidimensional, and parallel operations \cite{McMahon2023, Zhang2021, Brunner}. These systems promise to deliver high bandwidth, low latency processing and low power consumption \cite{7805240}. There have been several implementations of feed-forward neural networks (FF-NN) using free-space optics with spatial light modulators and diffractive elements \cite{Lin2018,Zhou2021}. Moreover, the integration of photonic devices on-chip has enabled the development of FF-NN using wavelength division multiplexing (WDM) with microring resonators (MRRs) \cite{Zhang:22,Ohno2022} and coherent systems with Mach-Zehnder interferometers (MZIs) \cite{Bandyopadhyay2022, Shen2017}. However, a main challenge for these approaches arises during the training stage, where implementing the widely used backpropagation algorithm, requires very complex architectures, which are very difficult to scale \cite{Hughes2018, doi:10.1126/science.ade8450}.

Another interesting option is to use unconventional machine learning architectures that require fewer resources for training. Spiking neural networks have shown their potential using photonic systems to increase the efficiency of AI systems \cite{2024NewCE...414010R,Inagaki2021, 10.1088/2634-4386/ad4b5b}. Moreover, photonic reservoir computing has been experimentally demonstrated to solve time-dependent problems using recurrent systems with physical or virtual nodes \cite{Appeltant2011,Nakajima2021, Shen:23}. 

Extreme learning machines (ELMs) \cite{1380068, HUANG2006489} are also and appealing alternative to traditional neural networks by simplifying its training process. ELMs are a type of FF-NN in which only the output layer weights are trained, while the while the connections of the input and hidden layer are randomly assigned and remain fixed. The idea is to randomly increase the dimensionality of an input vector to a new feature space where the data can be linearly separated. This methodology not only accelerates the training phase but also reduces the overall computational cost. ELMs are particularly well-suited for their photonic implementation on devices and systems that can inherently map the input optical signal into a higher dimensional space. Experimental demonstration of photonic ELMs (PELMs) using bulky systems have been already published using free-space propagation \cite{Pierangeli:21, M.Valensise:22}, speckle patterns \cite{10.1109/ICASSP.2016.7472872} and frequency multiplexed fibers \cite{Lupo:21}. ELMs in photonic integrated circuits have been also demonstrated by processing the scattered light of gratings coupled to microresonators \cite{10.1063/5.0156189}. This approach reduces the size of the PELM but still requires an external camera to record the hidden outputs. Moreover, the state of the random transformation of the photonic circuit is fixed which reduces the versatility of the PELM.

Here, we report a programmable photonic extreme learning machine (PPELM) using a general-purpose programmable processor \cite{Perez2018, Bogaerts2020, Zhuang:15, Zhou2020}. The programmable processor consists of a mesh of photonic tunable couplers in a hexagonal topology. We demonstrate how data can be encoded in both amplitude and phase using the tunable units. The processor can also be configured to adapt the number of inputs and outputs, making it suitable for different complex computational tasks. Numerous random matrices can be implemented on-chip by randomly setting the states of its building blocks. Furthermore, the inclusion of integrated photodetectors in the system allows the implementation of the nonlinear function directly on-chip. The final layer is trained and tested digitally revealing excellent performance in different complex tasks. In particular, the PPELM is experimentally validated using the Smartlight processor of iPronics on three classification tasks: the recognition of headers in a bit pattern, the categorization of iris flowers specimens, and the authentication of banknotes. Finally, we leverage the programmability of the PPELM to introduce methods to increase the accuracy of the models by optimizing the random hidden layer using an evolutionary algorithm and applying an ensemble technique where different models are trained in parallel thanks to a wavelength division multiplexing approach.

\section{Results}\label{results}
\begin{figure}[h]
    \centering
    \includegraphics[width=1.0\linewidth]{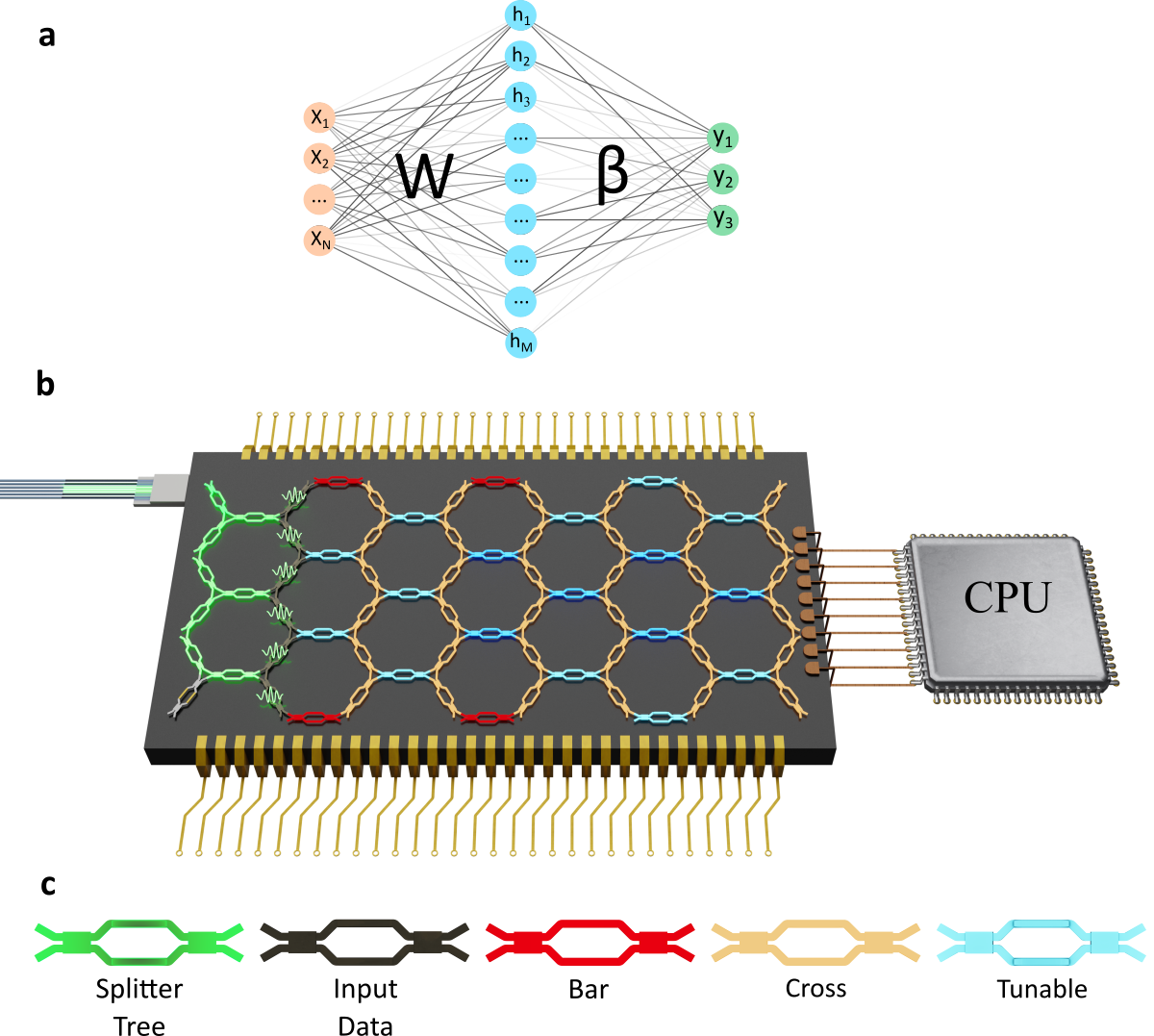}
    \caption{Programmable photonic extreme learning machines: \textbf{a} Diagram of the architecture of the extreme learning machines, \textbf{b} Schematic of the experimental setup for the implementation of programmable photonic extreme learning machines. The hexagonal mesh is programmed to multiply the input data by the random matrix and apply the nonlinear function using the square law of the photodetectors. The trainable weight matrix is multiplied on the CPU and, \textbf{c} different states of the PUCs on the hexagonal waveguide mesh.}
    \label{fig:hexagonal_mesh}
\end{figure}
\subsection{Extreme Learning Machines}
Extreme learning machines base their performance in the random projection of the input feature vector \textbf{X} that belongs to a space $\mathbb{R}^{N_{f}}$ into a higher-dimensional space $\mathbb{R}^{N_{H}}$ where the target classes can be linearly separated. $N_{f}$ refers to the number of input features and $N_{H}$ is the number of hidden nodes in the network. Mathematically, an input state \textbf{X} is first multiplied by a random matrix \textbf{W} and a set of biases \textbf{b} are added. To achieve a nonlinear random projection, an activation function \textit{f(.)} is applied:
\begin{equation}
    \textbf{H} = f(\textbf{W}\textbf{X} + \textbf{b})
\end{equation}
where \textbf{H} is the nonlinear random projection of the input state. 
The projected features are then linearly combined with a trainable weight matrix $\boldsymbol{\beta}$ to obtain the predicted outputs \textbf{O}.
Layer $\boldsymbol{\beta}$ can be trained using an analytical formula:
\begin{equation}
    \boldsymbol{\beta} = \boldsymbol{H^{\dag}T}
\label{eq:beta}
\end{equation}
where \textbf{T} is the ground-truth target matrix of the training dataset. 

\subsection{Programmable Photonic Circuit as an Extreme Learning Machine}\label{architecture}
We proposed the use of the Smartlight processor from IPronics Programmable Photonics \cite{Perez-Lopez2024} for the implementation of extreme learning machines in the photonic domain. Smartlight is a commercially available general-purpose programmable photonic integrated circuit with a hexagonal topology that can be tuned via software to create different photonic structures on the same platform. It comprises 72 programmable unit cells (balanced Mach-Zehnder interferometers) with 28 input-output optical ports. On-chip integrated photodetectors enable the conversion of the optical output power into the electronic domain through a readout layer. Further details on the characteristics of the chip and the hardware and software layers can be found in the Methods section.

A schematic of the programmable photonic extreme learning machine (PPELM) is shown in Fig. \ref{fig:hexagonal_mesh}. Light is coupled into the photonic integrated chip using an optical fiber array. The programmable unit cells (PUCs) in green are tuned to create a splitter tree that divides the light into four paths. The black PUCs, with a modulated wave on them, encode the input data into the amplitude of the light by changing their coupling ratio. Data has been previously normalized between -1 and 1, and the sign is encoded by changing the phase term of the PUCs, which is set to 0 for positive and $\pi$ for negative numbers. Additional information of the data encoding is presented in Appendix \ref{appendix1}. In our system, the maximum size of the input vector is 6, and we encode the features and biases. The remainder of the chip is programmed to implement a random transformation of the optical field. PUCs in red and gold are kept in bar and cross states respectively to avoid resonant structures inside the mesh when using a single-wavelength approach. As it will be explained later, the creation of resonators inside the mesh can be harnessed to increase the capabilities of our system when WDM is used. PUCs in blue are set as tunable, which means that their coupling ratio can be any number between $[0, 1]$ and the phase term is sampled from a uniform distribution between $[0, \pi]$. Once the input data is randomly projected, the nonlinear function is applied using the square law of the integrated photodetectors. Given the programmed structure light can be detected in up to 10 of the available photodetectors. During the experiments, we changed the state of the latest PUCs to use different numbers of outputs to measure the impact of the architecture's complexity. The measured currents were recorded using a CPU, where the training and application of the $\beta$ layer was carried out.

\subsection{PPELM on Classification Tasks}
We employed the PPELM to solve three different classification tasks: the IRIS flower and the banknote datasets classification tasks as well as header recognition task. For all three problems, 70$\%$ of the dataset was used for training and 30$\%$ for testing. The training of the models was performed as follows: firstly, we sampled the coupling ratios and phase terms of the tunable elements from a uniform distribution to create the random matrix. Secondly, data was normalized, encoded on the corresponding PUCs, and then multiplied by the random linear transformation. 

The nonlinear high-dimensional representation of the features was stored in the matrix $\textbf{H}_{train}$ of size $N \times N_{H}$, where $N$ is the number of points in the training dataset and $N_{H}$ is the number of hidden layer nodes. For each task, we repeated the experiments with 4, 6, 8, and 10 nodes. For every node count value, the random architecture was slightly modified to guide the light into the desired photodetectors and lose the minimum amount of light through non-monitored ports.
Following Eq. \ref{eq:beta}, the matrix $\textbf{H}_{train}$ was used to calculate the weights of the $\beta$ layer. Once the model was trained, we encoded the test set into the PPELM, multiplied it by the same random matrix, and obtained $\textbf{H}_{test}$. This matrix was multiplied by the final layer and the predicted outputs \textbf{O} were obtained. These results were compared with the expected targets \textbf{T} to measure the performance of the models. 
All trainings were repeated 40 times using different random projections. A summary of the obtained mean accuracies is presented in Table \ref{table:accuracy}.

\begin{figure}
    \centering
    \includegraphics[width=0.85\linewidth]{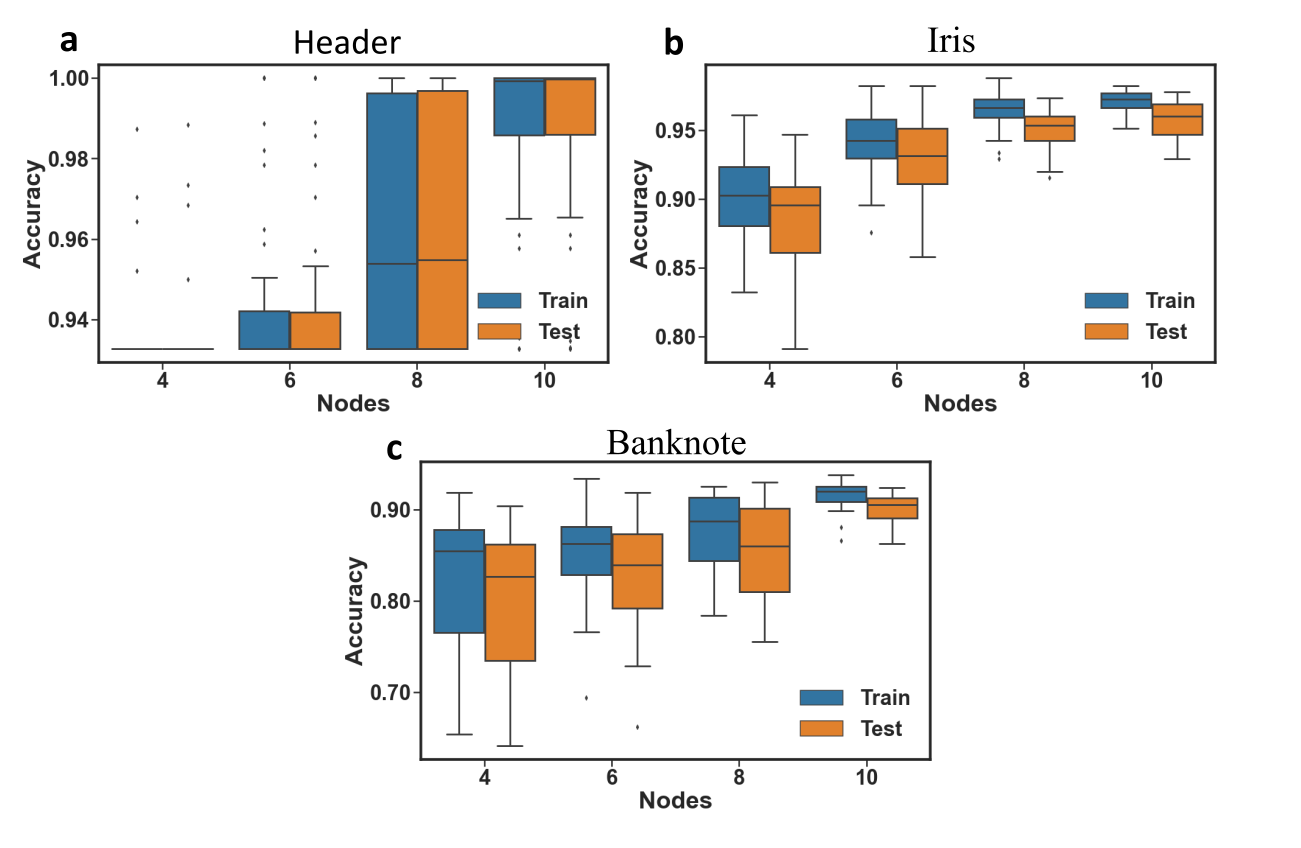}
    \caption{Accuracy of the of the PPELM in the \textbf{a} Header recognition task, the \textbf{b} Iris Flower and \textbf{c} Banknote Authentication Datasets Classification tasks when using 4, 6, 8 and 10 hidden nodes. In blue are the results of the train and in orange of the test set for 40 different random initializations.}
    \label{fig:graph_nodes}
\end{figure}

\begin{table}[h]
\caption{Test accuracy ($\%$) of the PPELM on three classification tasks with different number of hidden nodes}\label{table:accuracy}%
\begin{tabular}{@{}lllll@{}}
\toprule
Nodes & 4 & 6  & 8 & 10\\
\midrule
Header recognition    & 93.6$\pm$1.2 & 94.5$\pm$2.4 & 96.2$\pm$3.0 & 98.6$\pm$2.5\\
Iris    & 88.5$\pm$4.2   & 93.0$\pm$2.8  & 95.0$\pm$1.5 & 95.8$\pm$1.4  \\
Banknote    & 79.2$\pm$8.8  & 82.7$\pm$6.4  & 87.4$\pm$4.8 & 90.3$\pm$2.6  \\
\botrule
\end{tabular}
\end{table}

\subsubsection{Header recognition classification task}
The header classification task aims at recognizing a set of 4 bits within a data stream of 6000 points. We generated a random sequence of 1s and 0s, and classified each group of four bits as 1 if they matched the desired sequence, or 0 if they did not. In our case, we chose the sequence [1, 0, 0, 0] as the target. The results for the different number of nodes are shown in Fig. \ref{fig:graph_nodes}\textbf{a} which presents a boxplot of the accuracy obtained from the 40 models that we ran. The boxes cover the range from the first to the third quartile, with a line at the median. The whiskers extend to the last data point within 1.5 times the distance between the first and the third quartile. We can see how the accuracy increases with the number of nodes. As this is a simple task, a 100$\%$ accuracy can be obtained with 8 and 10 nodes.    

\subsubsection{Iris Flower dataset classification task}
The objective of the Iris Flower dataset task is to classify a set of flower specimens into three different subspecies: setosa, versicolor, and virginica. It comprises 150 feature vectors with 4 distinct characteristics: sepal length, sepal width, petal length, and petal width. Fig. \ref{fig:graph_nodes}\textbf{b} reveals that the PPELM was able to succesfully perform this complex task with high accuracy. In particular, \ref{fig:graph_nodes}\textbf{b}, shows the accuracy obtained for each number of hidden nodes. This is a more complex task than the header recognition task, and we can see how the variance in the results is reduced as the number of nodes increases, with the mean accuracy converging to a value close to 97$\%$ which matches the performance of other PELM proposals with a higher total network node count \cite{10.1063/5.0156189, Lupo:21}.

\subsubsection{Banknote dataset classification task}
The banknote dataset is a collection of 1,372 instances with 4 attributes: variance, skewness, kurtosis, and entropy of the banknote images, aiming to distinguish between genuine and forged banknotes. The results for the different models are shown in Fig. \ref{fig:graph_nodes}\textbf{c}, where accuracies over 90$\%$ are achieved in this complex task. This problem illustrates the importance of the initial random matrix. We can see that with all the hidden node sizes, there are models with high accuracies, but the variance is higher as the number of hidden nodes decreases. In the following section, we will introduce a method to find the optimal initial random matrix, thereby reducing the necessary length of the hidden layer in the PPELM. Moreover, the obtained mean accuracy is slightly lower than that presented by other PELMs. This is a consequence of the reduced number of hidden nodes. However, we will show how we can update the presented PPELM to match the accuracies of more complex photonic systems. 

\subsection{Evolutionary PPELM}
The results obtained from the three classification tasks highlight a key challenge in extreme learning machines:: the variance in achieved accuracies due to the randomness in the hidden layer weight matrix. While this issue is inherent to ELM models, its consequences are particularly significant in photonic systems. One option to mitigate the variance in model accuracy is to increase the number of hidden neurons, which in turn increases complexity, power consumption, and accumulated losses in photonic ELMs, thereby reducing their scalability. An alternative to increase the size of the system is to use a differential evolution algorithm (DE-PPELM) to find the optimal initialization of the random linear transformation \cite{ZHU20051759}. The programmability of our hexagonal mesh allows for the direct application of this approach into our system by optimizing the applied phases on the tunable PUCs.

To apply the differential evolution algorithm, at first, a set of N phase vectors is chosen as the initial candidates:
\begin{equation}
    \boldsymbol{\Theta_{G}} = \{ \boldsymbol{\theta}_{i,G} | i = 1, ... N\}
\end{equation}
where $\boldsymbol{\theta}_{i,G}$ = [$\theta_{0}$, $\theta_{1}$, ..., $\theta_{N_{TPUCs}}$] corresponds to the ith array with the phase value for all the tunable PUCs during generation G. The total number of vectors $N$ is referred to as the population size. For each of these vectors, the peformance of the PPELM is evaluated using a cost function $CF$. In our case, the cost function $CF$ is 1 minus the accuracy of the model on a validation set. The validation set is an extra division of the dataset (train - val - test) that allows for monitoring the algorithm's performance during training while leaving the test set untouched. 
The second step of the algorithm is the mutation, where three of the current candidates are combined using a control factor F as follows:
\begin{equation}
    \boldsymbol{v}_{i, G+1} = \boldsymbol{\theta}_{a,G} + F(\boldsymbol{\theta}_{b,G} - \boldsymbol{\theta}_{c,G})
\end{equation}
where a, b and c are independent random indices.
By selecting elements from the candidate and mutant sets, a new set of trial vectors $\boldsymbol{u}_{i, G+1}$ is built. These elements are chosen by sampling a random number b $\in$ [0, 1]. If b is lower than a crossover factor CR then the vector is picked from the mutation set $\boldsymbol{V}_{G+1}$, otherwise the vector is chosen from the set of candidates $\boldsymbol{\Theta}_{G}$. 
The performance of the model with the trial vectors is evaluated and compared to that achieved by the previous candidates. If a trial vector $\boldsymbol{u}_{j, G+1}$ presents a better $CF$ than the candidate $\boldsymbol{\theta}_{j, G+1}$ then the the trial vector replaces the candidate in the new generation. Otherwise, the new generation remains unchanged. If the difference between the two cost functions is less than 0.001, then we compare the norm of the trainable matrix $\boldsymbol{\beta}$ for both models and choose the solution with the lower norm. This is done because a lower norm in the second layer is related to a better generalization of the model \cite{10.5555/2627435.2670313}.

\begin{figure}[H]
    \centering
    \includegraphics[width=0.6\linewidth]{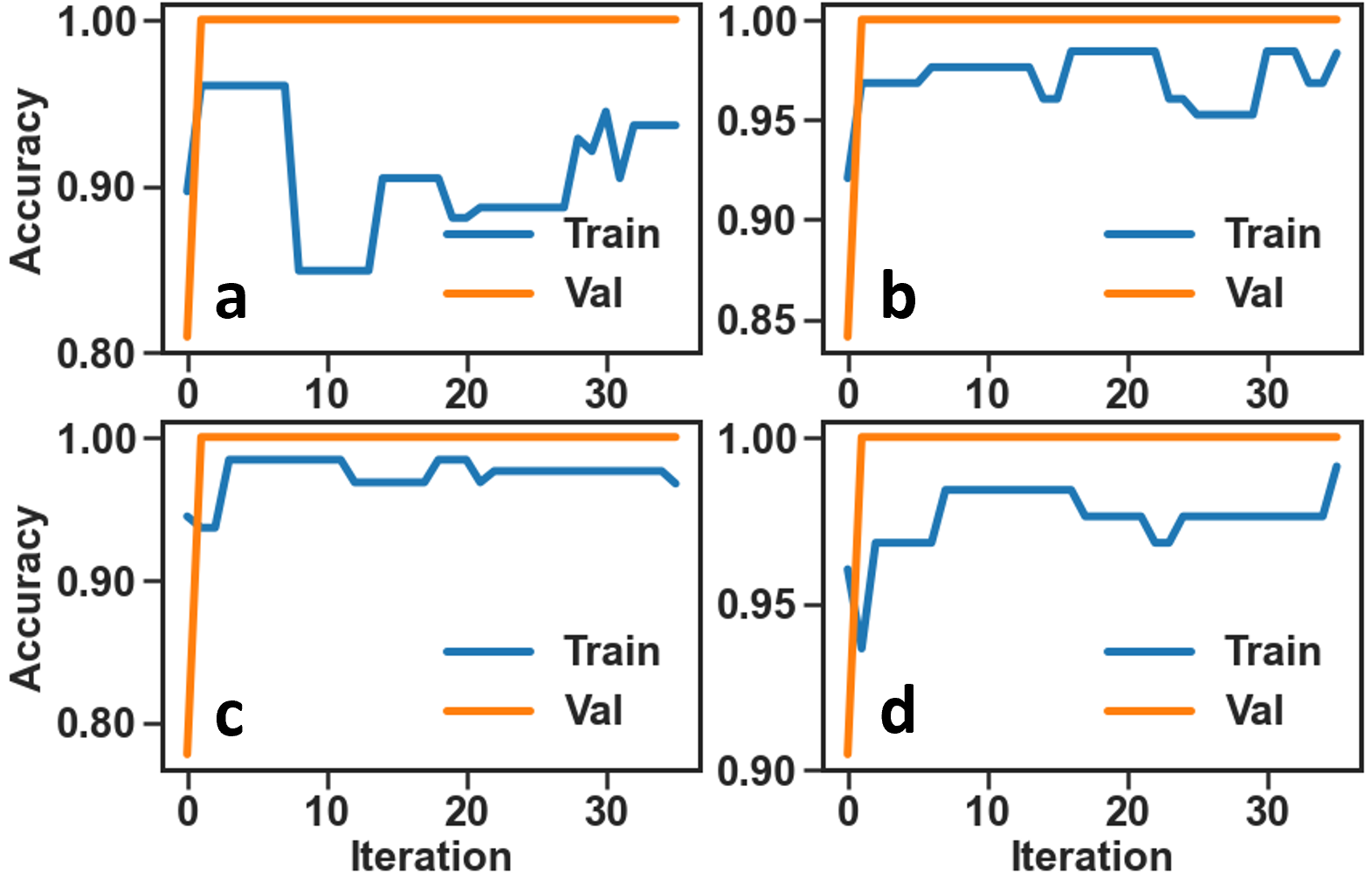}
    \caption{Evolution of the training and validation accuracy during 35 iterations of the DE-PPELM when using \textbf{a} 4, \textbf{b} 6, \textbf{c} 8, \textbf{d} 10 hidden nodes in the Iris Flower Dataset classification task.}
    \label{fig:DE-IRIS}
\end{figure}

\begin{figure}[H]
    \centering
    \includegraphics[width=0.6\linewidth]{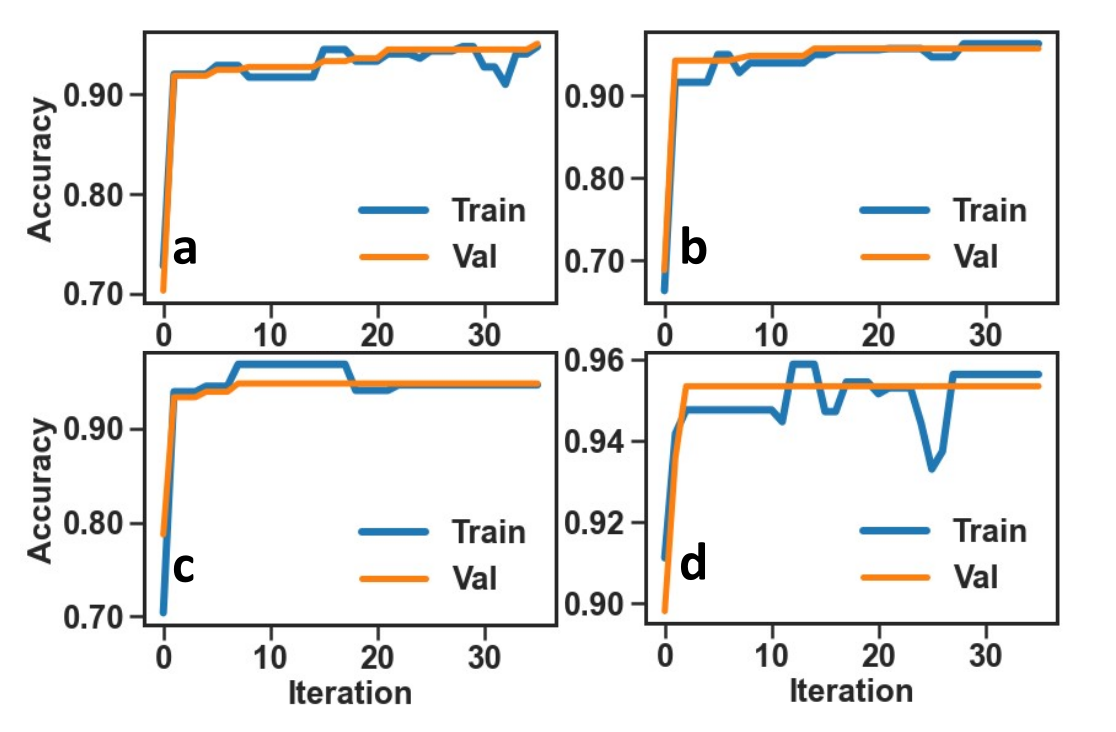}
    \caption{Evolution of the training and validation accuracy during 35 iterations of the DE-PPELM when using \textbf{a} 4, \textbf{b} 6, \textbf{c} 8, \textbf{d} 10 hidden nodes in the Banknote authentication dataset classification task.}
    \label{fig:DE-BANKNOTE}
\end{figure}

We evaluate the DE-PPELM on the Iris Flower and the Banknote Authentication dataset classification tasks. For this section we avoid the header recognition dataset as it presented very high accuracies for all the for all network hidden node counts investigated. As previously explained, for these experiments we divided the dataset into training, test and validation sets. We used 70 $\%$ of the dataset points for training, 15$\%$ for validation and 15$\%$ for testing. For each dataset, we initialised the random matrix using the differential evolution algorithm for all previous hidden node sizes. We used a population size of 10 for both tasks, a control factor F = 1 and a crossover term CR= 0.5. The DE-PPELM was trained during 35 iterations for all models. As the dataset sizes are small, especially the Iris Flower dataset, it is highly likely that high accuracies are achieved during the first iteration (which corresponds to 10 evaluations of the PPELM). To show how the algorithm can improve the accuracy of ill-conditioned random layers, we chose suboptimal candidates as a starting point during the first iteration.   

The evolution of the accuracies during the DE-PPELM training for the Iris flower dataset classification task is depicted in Fig. \ref{fig:DE-IRIS}. We can observe how the validation curve rapidly reaches a 100$\%$ accuracy, primarily due to the small size of the validation set. However, the algorithm continues to learn by attempting to reduce the norm of the output layer $\boldsymbol{\beta}$. Fluctuations in the training set are influenced by two factors. First, we aim to optimize the validation accuracy rather than the test accuracy, acting as a form of regularization for the training curve. Second, the small dataset size introduces higher divergences between the validation and training curves.
For the Banknote dataset classification tasks, the evolution of the accuracies is presented in Fig. \ref{fig:DE-BANKNOTE}. Here, we notice that, except for the 10-node model, the algorithm requires more steps to converge compared to the Iris task as a consequence of the larger dataset size. Additionally, in this case, the validation and training curves are more similar during the training of the algorithm. 

The final results after 35 iterations obtained on the test set are presented in Table \ref{table:DE-accuracy} for both classification tasks. For the Iris flower dataset classification task, we observe that for all hidden node sizes, the accuracy achieved on the test set is equal to or greater than the maximum accuracy obtained with the best model in the first experiment shown in Fig. \ref{fig:graph_nodes}. Regarding the Banknote task, we achieved accuracies exceeding 91$\%$ regardless of the number of hidden nodes. These results surpass the best results obtained during the first experiments in which a random hidden node weight matrix was set, hence demonstrating the pivotal contribution of the DE-PPELM algorithm for performance optimisation. Finally, it is essential to highlight how we were able to achieve high accuracies even with a reduced number of total network node count. This capability enables training highly complex computational tasks on a more compact photonic hardware than that required if the initial layer was directly sampled from a random distribution.

\begin{table}[h]
\caption{Test accuracy ($\%$) of the DE-PPELM on two classification tasks with different number of hidden nodes}\label{table:DE-accuracy}%
\begin{tabular}{@{}lllll@{}}
\toprule
Hidden Nodes & 4 & 6  & 8 & 10\\
\midrule
Iris    & 95.5   & 97.0  & 97.0 & 98.5  \\
Banknote    & 91.5  & 91.5  & 92.0 & 93.0  \\
\botrule
\end{tabular}
\end{table}

\subsection{WDM based Ensemble for increased accuracy}

\begin{figure}
    \centering
    \includegraphics[width=1.0\linewidth]{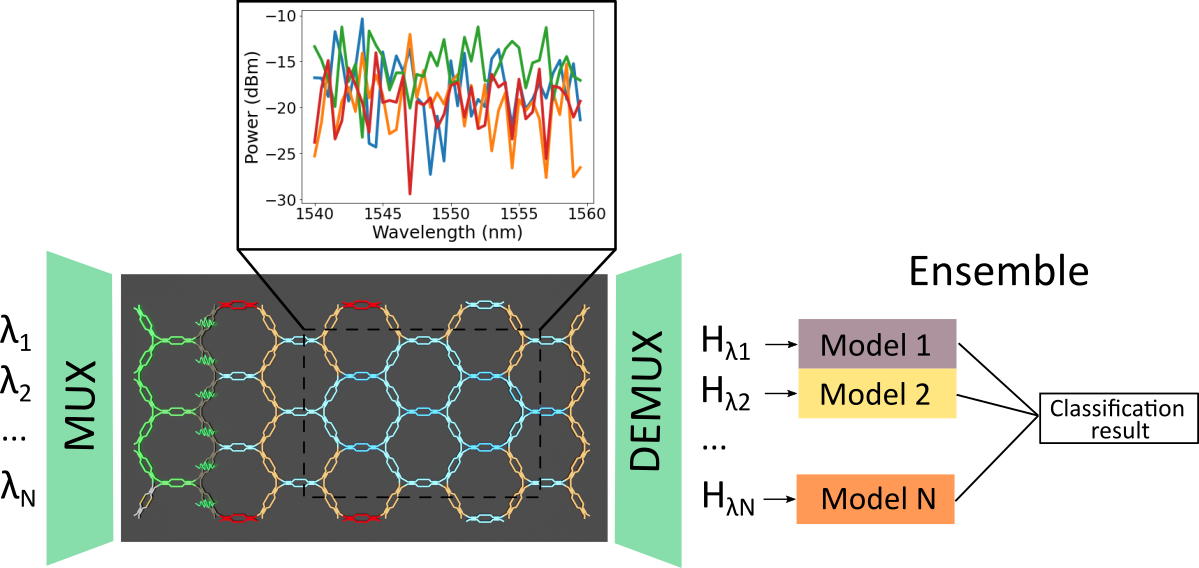}
    \caption{Ensembled extreme learning machines using WDM.}
    \label{fig:wdm_ensemble}
\end{figure}

In the previous section, we introduced an optimization algorithm to mitigate the impact of the random initialization of the internal layer of the ELM. Here, we propose an alternative approach to increase the accuracy of these systems by combining the predictions of a set of low accuracy models leveraging a widely studied technique called ensemble learning \cite{9893798}. In classical ELM approaches \cite{Abuassba2017, 5491079}, ensembling has demonstrated its ability to increase the efficacy in boosting accuracy on test sets while reducing overfitting. In our approach, we suggest taking advantage of the reconfigurability of our programmable photonic processor to train different models simultaneously in the photonic domain. Subsequently, we combine the results of the trained models to generate the final classification prediction. The proposed architecture uses WDM and is shown in Fig. \ref{fig:wdm_ensemble}. Using WDM to increase capacity by performing parallel computation has been proposed in the literature for photonic deep learning accelerators \cite{Dong2023, Feldmann2021, ou2024hypermultiplexed}. 

The architecture works as follows: Initially, different wavelengths are multiplexed into a single-mode fiber (SMF) and connected into the photonic chip using the same procedure as in our previous approaches. Secondly, both the training and testing sets are encoded on each of the wavelengths. Then, the random matrix architecture is modified by transitioning some of the diagonal PUCs from a cross to a tunable state. This adjustment facilitates the generation of resonant structures within the chip, thereby making the transfer function of the system wavelength-dependent. Following the random linear transformation, each output is demultiplexed, resulting in \textit{N} hidden output states corresponding to each of the \textit{N} input wavelengths. Subsequently, \textit{N} models are trained on a computer by computing the output matrix $\beta$. Each model undergoes training using a the same partition of the training and testing sets. Finally, the ensemble step involves combining the predictions of each model into a single array. This combined array is then utilized to compute the final prediction by training a final layer using Equation \ref{eq:beta}.

In our experiment, we substituted the MUX and DEMUX by a sequential process. We used a tunable laser and set one wavelength. We conducted the vector matrix multiplication and stored the results. This process was repeated with 40 wavelengths ranging from 1540 to 1560 nm, with a spacing of 0.5 nm between them. In Fig. \ref{fig:wdm_ensemble}, a graph with the resonant response of the system is illustrated. Each of the four represented coulours in the inset in Fig. \ref{fig:wdm_ensemble} corresponds to a different output of the mesh. The recorded results were used to derive the output matrix of each model, and the predicted outputs were aggregated for the final classification results. The results with varying numbers of used wavelengths are presented in Tables \ref{table:WDM-iris} and \ref{table:WDM-banknote}, where we compare the mean accuracy of the independent models with the ensemble results. Across all cases and classification tasks, the ensemble results consistently outperformed the mean accuracies during both training and testing. As in the evolutionary PPELM, the Iris flower classification task reaches its maximum accuracy with a low number of wavelengths as a consequence of the simplicity of the task. For the banknote classification task, it is important to highlight that we were able to obtain higher accuracies than the maximum level achieved with the random PPELM and with the evolutionary PPELM, with attained accuracies higher than 99 $\%$ when 25 or more wavelengths are used. 

\begin{table}[h]
\caption{Accuracy ($\%$) of the WDM-Ensemble on the Iris Flower classification task for different number of wavelengths}\label{table:WDM-iris}%
\begin{tabular}{@{}lllll@{}}
\toprule
$\#\lambda$ & Mean Train & Mean Test  & Ensemble Train & Ensemble Test \\
\midrule
5   & 95.7   & 93.4  & 96.4 & 96.4 \\
10    & 96.4  & 95.5  & 99.1 & 97.3\\
15    & 95.8  & 94.0  & 1.0 & 98.1 \\
20   & 95.3   & 93.3  & 1.0 & 98.1 \\
\botrule
\end{tabular}
\end{table}

\begin{table}[h]
\caption{Accuracy ($\%$) of the WDM-Ensemble on the Banknote classification task for different number of wavelengths}\label{table:WDM-banknote}%
\begin{tabular}{@{}lllll@{}}
\toprule
$\#\lambda$ & Mean Train & Mean Test  & Ensemble Train & Ensemble Test\\
\midrule
5   & 91.3   & 89.4  & 93.1 & 92.3   \\
10    & 88.8  & 87.4  & 95.3 & 93.1 \\
15    & 89.6  & 88.2  & 97.9 & 97.5  \\
20   & 89.1   & 87.8  & 99.1 & 99.1 \\
\botrule
\end{tabular}
\end{table}

\section{Discussion and Conclusions}\label{sec12}
We have demonstrated the implementation of photonic extreme learning machines (ELMs) using an integrated photonic programmable circuit. While previous photonic approaches relied on free-space propagation of light or application-specific photonic integrated circuits \cite{Pierangeli:21, 10.1109/ICASSP.2016.7472872, Lupo:21, 10.1063/5.0156189}, our solution is based on a general-purpose programmable photonic integrated platform.
To implement the programmable photonic ELMs, we first defined a splitter tree to divide the light into the number of dimensions of the input vector of the PPELM. Programming the splitter on-chip allowed us to maintain coherence inside the mesh, enabling us to exploit the complex-valued nature of the photonic devices in the integrated system. The normalized input vector was encoded in the amplitude of the light by adjusting the coupling of an array of PUCs, while the sign was encoded in the phase by simultaneously tuning both actuators of the PUC.

We incorporated the random layer of the ELMs by randomly selecting the coupling coefficient and phase terms of the tunable PUCs. We configured the state of the remaining PUCs to avoid resonant structures and ensure unidirectional light propagation. The nonlinear activation function was implemented using the square law of the integrated photodetectors. Finally, the output layer was calculated and digitally implemented using an analytic formula, similar to previous approaches. 

In an initial approach, we experimentally solved three classification tasks using an in-house created header recognition dataset, as well as the Iris flower and Banknote authentication datasets. We repeated all the experiments 20 times using network architectures with 4, 6, 8 and 10 hidden layer nodes. We showed how increasing the number of hidden nodes reduces the variability of the classification results that arises as a consequence of the random nature of the hidden layer. For the case of 10 hidden nodes we obtained mean accuracies of 98.6 $\pm$ 2.5, 95.8 $\pm$ 1.4 and 90.3 $\pm$ 2.6, for the Iris Flower and Banknote dataset classification tasks, respectively. Increasing the number of the number of hidden nodes may not always be the optimal solution as the scalability of photonic devices is limited and precision could be degraded \cite{10.1063/5.0070992}. As an approximation, the dynamic range at the outputs is:
\begin{equation}
    P_{in} - IL_{coupling} - N*IL_{PUC} - NF_{PD}
\end{equation}
where $P_{in}$ is the input power, $IL_{coupling}$ is the coupling instertion loss, $N$ is the minimum number of PUCs from the input to the photodetectors $IL_{PUC}$ are the insertion losses of the PUCs and $NF_{PD}$ is the noise floor of the photodetectors. Assuming $IL_{coupling}$ = 6 dB, $IL_{PUC}$ = 0.5 dB, $NF_{PD}$ = -35 dBm, $N$ = 15 and $P_{in}$ = 13 dBm the dynamic range is 34.5 dB.

As an alternative, we propose two solutions to enhance the classification accuracy of PELMs with a low number of hidden nodes. The first approach leverages the programmability of our system. We apply a differential evolution (DE) algorithm to optimize the hidden layer of the PELM. Our results demonstrate that this technique achieves accuracies equal to or better than the best values obtained for all network node sizes investigated, without the DE algorithm. Additionally, we find that fewer hidden nodes can be utilized with only a slight reduction in model accuracy. The main drawback of this solution is the increased computational resources and time required during the training of the evolutionary algorithm. Specifically, each iteration requires updating the state of the PUCs, which, in the case of thermo-optic phase shifters, is limited to microseconds. However, once the optimal random layer is determined and the model is trained, the weights remain fixed during inference, avoiding any additional overhead compared to the standard PPELM.

The second enhancement involves using an ensemble of models enabled by WDM techniques.. By manipulating part of the hexagonal mesh, we enable the creation of resonant structures with a wavelength-dependent transfer function. Multiple vector-matrix multiplications can then be performed simultaneously applying different wavelengths into the system. Although we conducted the experiment with one wavelength at a time, employing a MUX and DEMUX could facilitate calculating as many hidden outputs as wavelengths in parallel. The results from each wavelength model are then ensembled to derive a final classification prediction. This method yields promising results, surpassing those obtained in previous experiments. For instance, we achieved a 98.1$\%$ accuracy on the Iris classification task using 15 wavelengths and 8 outputs, and a 99.1$\%$ accuracy on the Banknote classification task using 20 wavelengths and 8 outputs. However, this approach requires additional photonic components, such as a MUX and DEMUX, for parallel computation. Although low-loss and high-bandwidth WDM filters have been demonstrated in recent literature using commercial foundries \cite{10526662}, the training of N models on the CPU may diminish the advantages of the PPELM.

In conclusion, we have shown how ELMs can benefit from the advantages of the photonic systems using an integrated programmable processor. Thanks to the latter's programmability we were able to test different random layers and proposed two methods to improve the performance of the models using and evolutionary algorithm and a WDM approach.

\section{Methods}\label{methods}
The Smartlight processor comprises a photonic chip manufactured in a standard 220-nm Silicon Photonics (SiPh) platform. The system integrates thermo-optical phase shifters with under-etched waveguides. Light input and output are facilitated using fiber arrays with edge coupling featuring a 127 $\mu m$ pitch. On-chip photodetection utilizes Germanium on silicon photodetectors. On-chip electrical pads are wirebonded to a PCB to enable the control of the phase actuators and the monitoring of the photodetectors. The photonic processor is controlled via software using the Smartlight python package provided by iPronics, allowing for full control of the state of each of the actuators.

For the single-wavelength and evolutionary PELM experiments, we utilized the ORTEL 10481S-FA laser with a fixed optical output power of 9.8 dBm operating at 1557 nm. For the WDM experiments, we employed the tunable laser TUNICS T100S-HP/CLU with a wavelength range from 1500 to 1680 nm and an output optical power of 13 dBm.

Training of the output layer of the PPELM was conducted on an Intel Core i5-10400H 2.6 GHz CPU with 16 GB of RAM.

\backmatter

\bmhead{Acknowledgements}
The authors would like to acknowledge Dr. Joshua Robertson and Dr. Giovanni Donati from the University of Strathclyde for their valuable discussions on photonic extreme learning machines and their contribution of valuable ideas to this work.

This work was supported by the H2020-ICT2019-2 Neoteric 871330 project, the European Research Council (ERC) Advanced Grant programme under grant agreement No. 101097092 (ANBIT), the ERC Starting Grant programme under grant agreement No. 101076175 (LS-Photonics Project), the EUR2022-134023 grant funded by CIN/AEI/10.13039/501100011033 and the European Union (NextGenerationEU/ PRTR), and the UKRI Turing AI Acceleration Fellowship Programme ‘PHOTON-AI’ (EP/V025198/1).

\section*{Declarations}

\begin{itemize}
\item Conflict of interest/Competing interests 

The authors have no conflicts to disclose
\item Data availability 

Data are available from the corresponding author upon reasonable request
\end{itemize}

\noindent

\begin{appendices}

\section{Data encoding in the Smartlight processor}\label{appendix1}
\begin{figure}[h]
    \centering
    \includegraphics[scale=0.65]{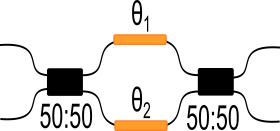}
    \caption{Programmable unit cell}
    \label{fig:bb_puc}
\end{figure}
Each feature is encoded on a programmable unit cell by changing the amplitude and phase of the optical signal. An schematic of the building block is presented in Fig. \ref{fig:bb_puc} and its transfer function is the following: 
\newcommand{\FF}{\vphantom{\frac{A_{i,1}y^2}{y^2}}}
\begin{equation}
\renewcommand{\arraystretch}{1.3}
    ie^{i\frac{\theta_{1} + \theta_{2}}{2}} 
    \begin{pmatrix}
        sin(\frac{\theta_{1} - \theta_{2}}{2}) & cos(\frac{\theta_{1} - \theta_{2}}{2})\\
        cos(\frac{\theta_{1} - \theta_{2}}{2}) & -sin(\frac{\theta_{1} - \theta_{2}}{2})
    \end{pmatrix}
    \label{eq:puc}
\end{equation}
The first step is to program the amplitude of the light. To do so, we need to know that the information will input the upper arm of the PUC and output the bottom arm, and thus, the output amplitude is:
\begin{equation}
    A_{out} = A_{in}cos(\frac{\theta_{1}-\theta_{2}}{2})
\end{equation}
Without losing generality, we can assume $A_{in}$ = 1 and $A_{out}$ = $|f|$, where $|.|$ is the modulus and $f$ is the feature value normalized between -1 and 1. Then:
\begin{equation}
    \Delta\Theta = \theta_{1}-\theta_{2} = 2*arccos(|f|)
    \label{eq:diff}
\end{equation}
Once the phase difference required for a certain amplitude is defined, then we need to include the output phase $angle(f)$ using the exponential term in \ref{eq:puc}:
\begin{equation}
    \theta_{1} + \theta_{2} = 2*angle(f)
    \label{eq:add}
\end{equation}
Combining \ref{eq:diff} and \ref{eq:add} we get the require phase values to encode a certain amplitude and phase:
\begin{gather}
    \theta_{1} = angle(f) + arccos(|f|) \\
    \theta_{2} = angle(f) - arccos(|f|)
\end{gather}




\end{appendices}


\bibliography{sn-bibliography}

\end{document}